\documentstyle[12pt]{article}
\newcommand{\be}{\begin{equation}}
\newcommand{\ee}{\end{equation}}
\newcommand{\ba}{\begin{eqnarray}}
\newcommand{\ea}{\end{eqnarray}}
\newcommand{\baa}{\begin{eqnarray*}}
\newcommand{\eaa}{\end{eqnarray*}}

\begin{document}
\thispagestyle{empty}

\vskip 2.0cm
{\renewcommand{\thefootnote}{\fnsymbol{footnote}}
\centerline{\large \bf Tertiary structure prediction of
C-peptide of}
\vspace{0.5cm}
\centerline{\large \bf ribonuclease A by multicanonical algorithm}

\vskip 2.0cm
 
\centerline{Ulrich H.E.~Hansmann
\footnote{\ \ e-mail: hansmann@ims.ac.jp}
 and Yuko Okamoto
\footnote{\ \ e-mail: okamotoy@ims.ac.jp}} 
\vskip 1.5cm
\centerline {{\it Department of Theoretical Studies}} 
\centerline{{\it Institute for Molecular Science}} 
\centerline {{\it Okazaki, Aichi 444, Japan}}

\medbreak
\vskip 3.5cm
 
\centerline{\bf ABSTRACT}
\vskip 0.3cm
 We have performed multicanonical Monte Carlo simulations of 
C-peptide of ribonuclease A.  It is known by CD and NMR experiments 
that this peptide has high $\alpha$-helix content in aqueous solution
and that the side-chain 
charges of residues Glu-2$^-$ and His-12$^+$ play an important role in 
the stability of the $\alpha$-helix.  
In order to confirm these experimental implications, we have
used two analogues of the peptide with charged and neutral side chains
of Glu-2 and His-12.
Two dielectric functions, distance-dependent and constant, are
considered to study the effects of solvent contributions.
All the simulations were started from random initial conformations.
Various thermodynamic quantities such as  
average helicity as a function 
of residue number and average distance between two side chains as
a function of temperature are calculated.
The results are found to be in accord with the implications of
CD and NMR experiments.  The lowest-energy conformation 
obtained has an $\alpha$-helix from Ala-4 to Gln-11 in complete
agreement with the corresponding structure deduced from an X-ray 
crystallography experiment of ribonuclease A.
It is shown that the 
salt bridge between the side chains of Glu-2$^-$ and Arg-10$^+$,
which is known to exist from both NMR and X-ray experiments,
is formed only when the side chains are properly charged. 
Its formation is greatly enhanced when the
distance-dependent dielectric function is used. 
\vfill
\newpage}
 \baselineskip=0.8cm
\noindent

The C-peptide, residues 1--13 of ribonuclease A, is known by CD and
NMR experiments
to have significant $\alpha$-helix formation in aqueous solution
at temperature near 0 $^{\circ}$C
\cite{Bz1,Bz2}.
In this article, we employ a multicanonical Monte Carlo simulation \cite{MU}
to study the $\alpha$-helix stability of C-peptide due to the side-chain
electrostatic interactions. 
The results are found to be in accord with various implications of the above
experiments.  
The lowest-energy conformation  
obtained by the simulation 
has an $\alpha$-helix from Ala-4 to Gln-11 in complete
agreement with the corresponding structure deduced from an X-ray 
crystallographic experiment of the whole ribonuclease A \cite{Xray}.
It is shown
that the characteristic salt bridge between Glu-2$^-$ and Arg-10$^+$,
which is known to exist both in the NMR experiment \cite{Bz2}
and in the X-ray experiment \cite{Xray}, is formed with significant
probability only when 
the side chains are properly charged and 
some solvation effects are included. 

The CD experiment of C-peptide showed that the side-chain 
charges of residues Glu-2$^-$ and His-12$^+$ play an important role in 
the stability of the $\alpha$-helix, while the rest of the charges of other
side chains do not \cite{Bz1}. 
A previous simulation work \cite{KONF} by Monte Carlo simulated annealing 
\cite{SA} confirmed the $\alpha$-helix formation
and the importance of the electrostatic interactions 
of the above two side chains for the stability of the helix.  
The simulation was performed in gas phase, however, and it failed in
obtaining the characteristic salt bridge between Glu-2$^-$ and Arg-10$^+$.
    
In this work, we 
used two analogues of C-peptide in order to study
the importance for $\alpha$-helix stability due to the electrostatic 
interactions of the side-chain
charges of residues Glu-2 and His-12.
The amino-acid sequences of these analogues 
are 
K$^+$E$^-$TAAAK$^+$FER$^+$QH$^+$M and
K$^+$ETAAAK$^+$FE$^-$R$^+$QHM.  
We refer to the former as Peptide I
and the latter Peptide II hereafter.  The main difference between the
two peptides is the charges of residues Glu-2 and His-12
(both are charged for Peptide I and neutral for Peptide II, respectively).
The potential energy function
that we used is given by the sum of
the electrostatic term, 12-6 Lennard-Jones term, and
hydrogen-bond term for all pairs of atoms in the peptide together 
with the torsion term for all torsion angles.
The energy parameters were adopted
from ECEPP/2 \cite{EC3}.
A distance-dependent dielectric function \cite{O} was used to mimic
the presence of water. A constant dielectric function ($\epsilon = 2$)
was also used for a comparison with gas-phase simulations.
The computer code KONF90 \cite{KONF} was used. 

The Monte Carlo method that we used is multicanonical algorithm \cite{MU},
which belongs to a class of {\it generalized-ensemble algorithms} (for a
discussion and comparison of these algorithms, see, for instance,
Ref.~\cite{HO96}).
This method was introduced to the protein folding problem a few years
ago \cite{HO}, and the effectiveness of the method has been established
for oligopeptide systems \cite{HO94,HO95}.
The method allows one to sample a wide range of configuration space and 
overcome the multiple-minima problem that is responsible for the very
long equilibrization time required by conventional methods.
>From a single simulation run, it thus enables one to obtain not only
the lowest-energy conformation but also any thermodynamic quantity 
over a wide range of temperatures \cite{MU,HO}. 
In the present work, we performed four multicanonical Monte Carlo simulations
of 1,000,000 Monte Carlo sweeps each, where one Monte Carlo sweep updates
all the torsion angles in the peptide once.  The four runs are: one
with distance-dependent dielectric function for Peptide I, one with
constant dielectric function for Peptide I, and two corresponding runs
for Peptide II.
The simulations were started from completely random initial conformations.

We first examine how much $\alpha$-helix formation we obtain 
by the simulations.  We found that on the average 65.2 (1.0) \% of the
residues are in $\alpha$-helix state 
at temperature $T = 273$ K
for Peptide I with distance-dependent dielectric function, while 
the value was 52.3 (3.5) \% for Peptide II with the same dielectric function
(the numbers in parentheses are errors).  
Here, a residue is defined
to be in $\alpha$-helix state if the backbone dihedral angles 
$( \phi, \psi)$ fall in the range
$( -70 \pm 30^{\circ}, -37 \pm 30^{\circ})$. 
The average length of $\alpha$-helix at this temperature are, likewise,
7.7 (2.2) residues long and 4.8 (0.5) residues long 
for Peptide I and Peptide II,
respectively.  Furthermore, at this temperature the average energy differences 
between helical conformations and non-helical conformations,
$\Delta E~(\equiv <E_H> - <E_C>)$, are
$-20.5$ (3.5) and $-3.1$ (4.7) for Peptide I and Peptide II, respectively.
Here, $<E_H>$ and $<E_C>$ stand for
the average total potential energies of helical conformations and non-helical
conformations, respectively (and a helical conformation is defined to be
a conformation that has at least
3 successive residues in the $\alpha$-helix state). 
The large difference in $\Delta E$ implies that a helical conformation
is energetically favored.
All these results support the experimental fact that
the side-chain charges of the residues Glu-2 and His-12 enhance the
$\alpha$-helix stability of C-peptide \cite{Bz1}.  

In Figure~1 we compare the average
\% helicity of the two peptides at $T = 273$ K as a function 
of residue number obtained by simulations with the distance-dependent
dielectric function.  The overall helicity is larger in Peptide I than
in Peptide II as just discussed above.  The helicity is very high
from residue 4 to residue 11 and it is very low from residue 1 to residue 3
(and residues 12 and 13) for Peptide I.  This is in accord with the
implications of the NMR experiment of C-peptide, where they found lowered
population of helices for residues 1--3 and high helix content
for residues 5--12 \cite{Bz2}.  The results for Peptide II, on the other
hand, is inconsistent with the NMR data in that they predict high helicity
for residue 2 and very low helicity for residue 7.  These findings again
support the fact that the 
charges of the residues Glu-2 and His-12 are important for the
$\alpha$-helix stability of C-peptide.

The lowest-energy conformations of the two peptides obtained by the
simulations with the distance-dependent dielectric function are now
compared.  In Table~I we give the backbone dihedral angles of these
conformations together with those of a structure deduced by the
X-ray experiment \cite{Xray}.  The conformation of Peptide I has an
$\alpha$-helix in residues 4--11 in complete agreement with the
X-ray data, while that of Peptide II has an extended $\alpha$-helix
only for residues 8--12.  The root-mean-square (r.m.s.) deviations of 
these conformations from the X-ray structure are also presented in the
Table.  One finds that the backbone structure for Peptide I is 
very similar 
to that of the X-ray data (r.m.s. distance of 1.4 \AA).

The lowest-energy conformation of Peptide I is shown in Figure~2A.
>From the Figure we see that
the characteristic salt bridge between Glu 2$^-$ and Arg 10$^+$,
which exists both in the NMR data \cite{Bz2}
and in the X-ray data \cite{Xray}, is indeed formed. 
In Figure~2B this structure and the X-ray structure are
displayed together in a superposition.  One can see that the two 
tertiary structures
are quite similar to each other 
(r.m.s. distance is 2.7 \AA~ from Table~I).

The formation of the salt bridge can be studied by calculating
the average distance between the side chains of Glu-2 and Arg-10
as a function of temperature.  Here, the distance between these side
chains is defined to be the smallest of the distance between
O$^{\epsilon}$ of Glu-2 and H$^{\eta}$ of Arg-10.  The results for all
four runs are given in Figure~3.  From the Figure one finds that
the results for Peptide I with distance-dependent dielectric
function give the shortest average distance between the two
side chains at low temperatures.
This implies that the salt bridge between Glu-2 and Arg-10
is favored most when 
the side chains of residues Glu-2 and His-12 are charged (Peptide I
rather than Peptide II) and 
some solvation effects are included (distance-dependent dielectric
function rather than constant one). 

In this article, we have presented the results of multicanonical
Monte Carlo simulations applied to the tertiary-structure prediction 
of C-peptide of ribonuclease A.  The results were in good agreement
with various implications of CD, NMR, and X-ray experiments.
It should be emphasized that the
simulations were performed from completely random initial conformations
and that no structural information from experiments was used as input.
Furthermore, it is a great advantage of multicanonical algorithm
over other methods
that one needs only a {\it single} simulation run to obtain any thermodynamic
quantity for a wide range of temperatures.

\vspace{0.5cm}
\noindent
{\bf Acknowledgments}: \\
Our simulations were  performed on computers in the Computer Center 
of the Institute for Molecular Science (IMS), Okazaki,
Japan.
This work is supported by a Grant-in-Aid for Scientific Research from the
Japanese Ministry of Education, Science, Sports and Culture.


\newpage
~~\\

\noindent
{\bf Table~I.}  Backbone dihedral angles (in degrees)$^{\rm a}$
of the lowest-energy conformations of \\
\hspace*{1.8cm} Peptides I and II for distance-dependent dielectric function 
obtained from \\
\hspace*{1.8cm} the multicanonical 
simulations and those deduced from the X-ray data \cite{Xray}, \\
\hspace*{1.8cm} together with the r.m.s. deviations (in \AA) 
from the X-ray structure.$^{\rm b}$ \\ 
\begin{center}
\begin{tabular}{ccccccccc} \hline
\rule{0mm}{3mm} \\
 &\multicolumn{2}{c}{X-ray}&~~~&
\multicolumn{2}{c}{Peptide I}&~~~&
\multicolumn{2}{c}{Peptide II} \\
\rule{0mm}{3mm} \\
\hline
\rule{0mm}{3mm} \\
Residue&$\phi$&$\psi$& &$\phi$&$\psi$& &$\phi$&$\psi$ \\
\rule{0mm}{1mm} \\
Lys-1 &    & 175 & & $-7$ & $-51$ & & $-2$ & $-41$ \\
Glu-2 & $-58$ & 136 & & $-83$ & 79 & & $-60*$ & $-34*$ \\
Thr-3 & $-68$ & 159 & & $-102$ & 156 & & $-79$ & 69 \\
Ala-4 & $-59*$ & $-45*$ & & $-57*$ & $-32*$ & & $-69$ & 125 \\
Ala-5 & $-64*$ & $-48*$ & & $-70*$ & $-49*$ & & $-159$ & 166 \\
Ala-6 & $-64*$ & $-34*$ & & $-59*$ & $-37*$ & & $-57*$ & $-41*$ \\
Lys-7 & $-63*$ & $-42*$ & & $-64*$ & $-52*$ & & $-62$ & 116 \\
Phe-8 & $-61*$ & $-42*$ & & $-61*$ & $-36*$ & & $-55*$ & $-38*$ \\
Glu-9 & $-58*$ & $-46*$ & & $-63*$ & $-48*$ & & $-55*$ & $-47*$ \\
Arg-10 & $-64*$ & $-37*$ & & $-62*$ & $-36*$ & & $-75*$ & $-33*$ \\
Gln-11 & $-71*$ & $-28*$ & & $-72*$ & $-37*$ & & $-57*$ & $-40*$ \\
His-12 & $-120$ & $-12$ & & $-172$ & 120 & & $-86*$ & $-41*$ \\
Met-13 & $-104$ & 130 & & $-60$ & 133 & & $-106$ & 98 \\
\rule{0mm}{3mm} \\
\hline
\rule{0mm}{3mm} \\
\multicolumn{5}{l}{R.m.s. deviation from X-ray structure}& & & &  \\
\rule{0mm}{1mm} \\
Backbone &\multicolumn{2}{c}{0.0}&~~~&
\multicolumn{2}{c}{1.4}&~~~&
\multicolumn{2}{c}{5.2} \\
All Atoms &\multicolumn{2}{c}{0.0}&~~~&
\multicolumn{2}{c}{2.7}&~~~&
\multicolumn{2}{c}{6.1} \\
\rule{0mm}{3mm} \\
\hline
\end{tabular}
\end{center}
~~\\
\noindent
$^{\rm a}$ The asterisks indicate that the corresponding
residues are in the $\alpha$-helix state, where a residue is defined
to be in $\alpha$-helix state if the dihedral angles 
$( \phi, \psi)$ fall in the range
$( -70 \pm 30^{\circ}, -37 \pm 30^{\circ})$. \\
$^{\rm b}$ The X-ray structure was taken from the Brookhaven Protein Data
Bank file 8RAT \cite{Xray}.  The r.m.s. distance was calculated with respect to 
non-hydrogen atoms only. \\  
        
\newpage
\centerline{\bf Figure Legends}

\begin{itemize}
\item Figure~1: Average \% helicity of C-peptide analogues, Peptide I (PI)
      and Peptide II (PII), at $T=273$ K as a function of residue number.
      The results are for the distance-dependent dielectric function.
      Here, a residue is defined
      to be in $\alpha$-helix state if the backbone dihedral angles 
      $( \phi, \psi)$ fall in the range
      $( -70 \pm 30^{\circ}, -37 \pm 30^{\circ})$.
      Each result was obtained from a multicanonical
      simulation of 1,000,000 Monte Carlo sweeps.
\item Figure~2: (A) The lowest-energy conformation of 
      C-peptide of ribonuclease A (Peptide I) obtained by a multicanonical
      simulation of 1,000,000 Monte Carlo sweeps with distance-dependent 
      dielectric function.  The side-chain atoms are suppressed except 
      for those of Glu-2$^-$ and Arg-10$^+$ that form a salt bridge.
      These side chains are labeled in the figure.  
      The N terminus and
      the C terminus are also labeled by N and C, respectively.
      The figure was created with Molscript \cite{Molscript}.
      (B) The lowest-energy conformation of Figure~2A (black sticks) 
      and the corresponding X-ray structure (gray sticks) \cite{Xray}
      superposed.  All the atoms in the backbone and 
      side chains are shown here,
      but the hydrogen atoms are suppressed.
      The N terminus and the C terminus are labeled by N and C, respectively.
      The figure was created with RasMol \cite{Rasmol}.
\item Figure~3: Average distance $<d(2-10)>$ (in \AA) between the side chains 
      of Glu-2 and Arg-10 as a function of temperature (in K).  Here, 
      the distance $d(2-10)$ is defined to be the smallest of the distance 
      between O$^{\epsilon}$ of Glu-2 and H$^{\eta}$ of Arg-10.  
      PI and PII correspond to Peptide I and Peptide II, respectively.
      ${\rm epsi}={\rm dis}$ and ${\rm epsi}=2$ stand for distance-dependent
      dielectric function and constant dielectric function, respectively. 
      Each result was obtained from a multicanonical 
      simulation of 1,000,000 Monte Carlo sweeps.
\end{itemize}
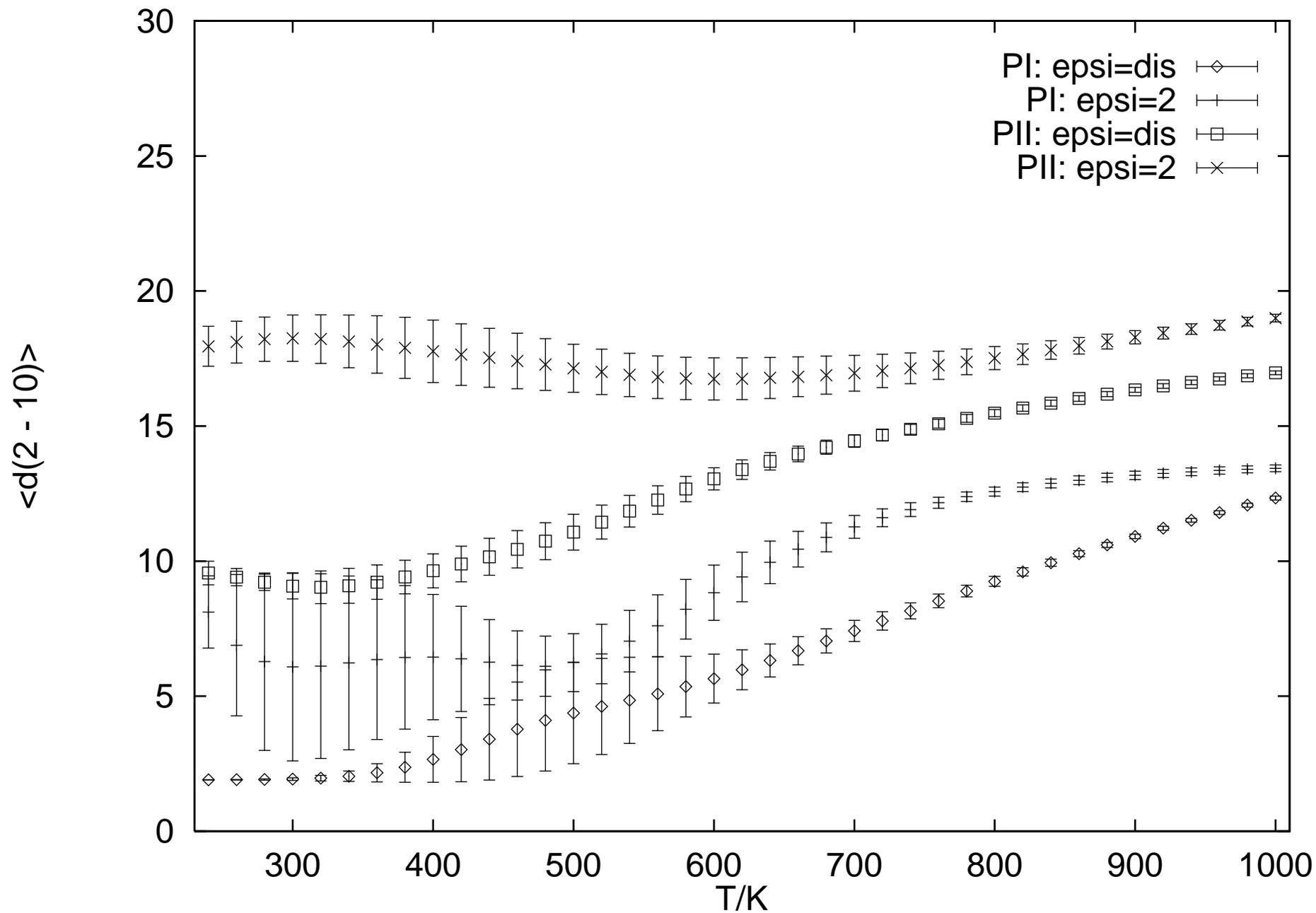
\end{document}